\documentclass[12pt,letterpaper]{article}

\usepackage{amsmath}
\usepackage{amsfonts}
\usepackage{amssymb}
\usepackage{graphicx}

\renewcommand\({\left(}
\renewcommand\){\right)}
\renewcommand\[{\left[}
\renewcommand\]{\right]}

\let\d=\partial
\let\z=\zeta

\newcommand\rd{{\rm{d}}}

\newcommand\ee{\end{equation}}

\newcommand\be{\begin{equation}}

\newcommand\eea{\end{eqnarray}}

\newcommand\bea{\begin{eqnarray}}

\def\d{\partial}

\begin{document}

\vspace*{1cm}

\begin{center}

\def\thefootnote{\fnsymbol{footnote}}
{\Large \bf Action approach to cosmological perturbations:}\\
\vspace{0.3cm} {\Large \bf the 2nd order metric in matter
dominance}
\\[0.5cm]
{\large Lotfi Boubekeur$^{\rm a}$, Paolo Creminelli$^{\rm a}$,\\[.1cm]
 Jorge Nore\~na$^{\rm b}$,
and Filippo Vernizzi$^{\rm a}$}
\\[0.5cm]

{\small \textit{$^{\rm a}$ Abdus Salam ICTP, Strada Costiera 11,
34014 Trieste, Italy}}

\vspace{.2cm}

{\small \textit{$^{\rm b}$ SISSA/ISAS, Via Beirut 2, 34014
Trieste, Italy}}

\end{center}

\vspace{.8cm}

\hrule \vspace{0.3cm}
{\small  \noindent \textbf{Abstract} \\[0.3cm]
\noindent We study nonlinear cosmological perturbations during the
post-inflationary evolution, using the equivalence between a
perfect barotropic fluid and a derivatively coupled scalar field
with Lagrangian $[-(\partial \phi)^2]^{(1+w)/2w}$. Since this Lagrangian
is just a special case of {\em k}-inflation, this approach is
analogous to the one employed in the study of non-Gaussianities from
inflation. We use this method to derive the second order metric
during matter dominance in the comoving gauge directly as function
of the primordial inflationary perturbation $\zeta$. Going to
Poisson gauge, we recover the metric previously derived in the
literature.

\vspace{0.5cm}  \hrule
\def\thefootnote{\arabic{footnote}}
\setcounter{footnote}{0} \vspace{0.5cm}

\section{Introduction}

Recently, there has been an extraordinary improvement in the
accuracy of cosmological observations, especially regarding the
statistical properties and the evolution of perturbations around a FLRW Universe.
In order to fully exploit the data, it has become necessary to go
beyond linear perturbation theory. This is mandatory, for
instance, if one wants to study primordial non-Gaussianities which, by
definition, are sensitive to the non-linear interactions of the theory.

The calculation of primordial non-Gaussianities in
various models is based on a perturbative expansion of the action
-- or of the equations of motion -- of one or more scalar fields. In
general, the mixing of these scalar fields with gravity cannot be
neglected. 
This can be done by using an ADM approach
which allows to write an action for the relevant degrees of
freedom taking into account gravitational perturbations
\cite{Maldacena:2002vr}. In this paper we want to extend this
formalism to the study of cosmological perturbations {\em after}
inflation, during the standard FLRW evolution.

During its evolution, the Universe is filled with cosmological fluids
such as radiation or cold dark matter. At first sight a fluid is very 
different from a scalar field, so that it is not clear
how to extend the aforementioned approach to our case. However, in
most cases the cosmological fluids can be considered as perfect
and barotropic. In this case, as we will review in
section~\ref{fluid}, the dynamics of the fluid is exactly
equivalent to the one of a derivatively coupled scalar field,
playing the role of the fluid potential velocity. Using this
equivalence, the usual treatment of perturbations in the primordial Universe,
used when they are generated and stretched out of the Hubble radius,
can be extended to their subsequent evolution in the FLRW
Universe, when the modes re-enter the Hubble radius.

In this article this approach is applied to find the exact metric
at second order during the matter dominated era. Our
results are compatible with those obtained previously in
\cite{Matarrese:1997ay,Bartolo:2005kv}, by directly working at the level
of the Einstein equations. The knowledge of the
second order metric is the first step for the calculation of any
observable beyond the linear approximation. This calculations are
quite relevant: in the case of the CMB
anisotropy the expected magnitude of a generic second order effect is
comparable to the sensitivity of the forthcoming Planck satellite.

As we will see,
the calculation of the second order metric will closely parallel the one of the three-point function
of scalar and tensor perturbations generated during inflation. In
particular, we will make use of the action 
calculated for this purpose for {\em k}-inflation
\cite{Seery:2005wm,Chen:2006nt}, expanded up to third order.
In this context, it is natural to use the comoving curvature
perturbation $\zeta$ as the variable describing the
scalar perturbations, the same which is
commonly employed for the quantization of scalar
fluctuations during inflation. This variable is also well-known to
be nonlinearly conserved on super-Hubble scales independently of
the details of the cosmological evolution \cite{Salopek:1990jq}.
For this reason it is the natural variable to set the initial
conditions for the post-inflationary dynamics.
Therefore, an advantage of our approach is that the dynamics of
perturbations is described by the same variable and the same approach, from its
generation at early times to the late times observations.

As a warm-up, we start with linear perturbation theory in
section~\ref{bd}. Then, in section~\ref{second}, we extend our
calculation to second order for scalar and tensor modes, assuming
that primordial gravitational waves are negligible. In this way we
obtain the full second order metric in the comoving gauge including
tensor modes generated by scalars. In section \ref{poisson} we
check our results with the ones obtained in Poisson gauge
\cite{Matarrese:1997ay,Bartolo:2005kv} after a suitable second
order gauge transformation. Conclusions are drawn in section \ref{conclusions}.

\section{A fluid as a scalar field} \label{fluid}
In cosmology, when dissipative phenomena
are negligible, the energy content of the Universe can be
approximated as a sum of perfect fluids. A perfect fluid is defined to have a stress-energy
tensor of the form
\be
T_{\mu\nu} = (\rho+ p ) u_\mu u_\nu + p g_{\mu\nu} \;, \label{set_pf}
\ee
where $\rho$ and $p$ are the energy density and the pressure, while
$u^\mu$ is the fluid 4-velocity.
Cosmological fluids are usually taken
to be irrotational. This assumption is
justified by the absence of vorticity in the initial conditions set
by inflation and by the fact that vorticity is diluted by the
expansion of the Universe. The fluids are also taken to be barotropic,
i.e., their pressure is a function of the energy density only, $p =
p(\rho)$. Under these conditions, each fluid is characterized by
a single scalar function, so that it is not surprising that its
dynamics can be described in terms of a scalar field.

Indeed, let us consider a derivatively coupled scalar  $\phi$ in
Minkowski spacetime with Lagrangian density\footnote{For Lagrangian
  approaches which describe also vorticous motion and
  non-barotropic perfect fluids see \cite{Taub,Schutz:1970my,Dubovsky:2005xd}.}
\be
{\cal L} = P(X)\;, \qquad X \equiv -\partial_\mu \phi \partial^\mu \phi \;. \label{lphi}
\ee
Varying the action yields the equation of motion
\be
\partial_\mu [P'(X)\partial^\mu \phi ] =0\;. \label{eom2}
\ee
The stress-energy tensor of this field is given by
\be
T_{\mu \nu} =  2P'(X) \partial_\mu \phi \partial_\nu \phi + P(X)
g_{\mu \nu} \;, \label{set_phi}
\ee
which is of the perfect fluid form (\ref{set_pf}) if we identify
\be
\rho = 2P'X -P\;, \qquad p = P\;, \qquad u_\mu = \frac{\partial_\mu
  \phi}{\sqrt{X}}\;. \label{fq}
\ee
The perfect fluid interpretation makes sense only if $\partial_\mu
\phi$ is everywhere timelike and future directed.
Projecting the conservation equation of the stress-energy tensor of a
fluid, $\partial_\mu T^\mu_\nu=0$, along and orthogonal to the fluid
flux yields the equation of conservation of energy and the Euler equation.
In the case of the stress-energy tensor (\ref{set_phi}), the Euler
equation is a trivial identity, while the conservation of the energy
is equivalent to the equation of motion (\ref{eom2}).

The equation of motion (\ref{eom2}) can be interpreted as the
conservation of the current 
\be
J^\mu = 2 \sqrt{X} P'(X) \cdot u^\mu \;.
\ee 
This conservation is a consequence (by
N\"other theorem) of the invariance of the action under
shift of $\phi:$ $\phi \to \phi + {\rm const}$. From the fluid point of view
this current describes the conserved particle density flux $J^\mu \equiv n
u^\mu$, where $n$ is the number particle density. Therefore one can identify $n = 2 \sqrt{X} P'(X)$. This yields
a physical interpretation of the norm of $\partial_\mu \phi$:
$\sqrt{X} = (\rho+P)/n$ is the so called specific
inertial mass \cite{Taub,Schutz:1970my}. It is also
straightforward to verify that
\be
\sqrt{X} = \frac{d\rho}{dn} \;,
\ee
so that $\sqrt{X}$ acts as a sort of conjugate variable
with respect to $n$ \cite{Dubovsky:2005xd}.

It is well known that for a perfect fluid the entropy per particle is conserved
along the fluid flow. This can be checked using the continuity
equation $\partial_\mu J^\mu=0$, i.e. eq.~(\ref{eom2}), and the
energy conservation $u^\mu \partial_\nu T^\nu_\mu=0$
\cite{Weinberg}.
Furthermore, as we have discussed, our approach describes a barotropic
fluid which implies that the entropy per particle is everywhere
constant. In other words, the Lagrangian (\ref{lphi})
can only describe mechanical excitations of the fluid. It
cannot take into account dissipative irreversible processes like heat
conduction or viscosity.

We will be interested in studying perturbations around a homogeneous
configuration. In Min\-kow\-ski spacetime $\phi=ct$ is a solution of
the equation of motion for any $c$. Different values of $c$ describe
different unperturbed values of the energy density $\rho(c^2)$. The
dynamics of fluctuations around this background can be studied by
expanding the Lagrangian using $\phi = ct+\delta\phi(t,\vec x)$, where $\delta\phi$
describes the compressional mode of the fluid.
At second order we obtain
\be
{\cal L} = P'(c^2) [\dot {\delta\phi}^2  - (\nabla \delta\phi)^2 ] + 2 P''(c^2) c^2
\dot {\delta\phi}^2 \;.
\ee
From this expression one sees that the speed of sound of the
excitations is given by
\be
c_s^2=\left. {P'(X)\over P'(X)+ 2 X P''(X)} \right|_{X=c^2}\;,
\label{cs}
\ee
which, as expected, is the usual adiabatic speed of sound in a
barotropic fluid,
\be
c_s^2 = \left. \frac{p'(X)}{\rho'(X)} \right|_{X=c^2}=
\frac{dp}{d\rho} \;.
\label{cs2}
\ee
Note that although the Lagrangian (\ref{lphi}) is Lorentz
invariant, this symmetry is spontaneously broken by the vacuum
$\phi=ct$. For this reason the speed of perturbations will in general differ
from the speed of light $c_s^2=1$.

A standard case of barotropic fluid is given by a linear equation of
state $p=w \rho$ with $w=$ const. From the first two equalities in
eq.~(\ref{fq}) one deduces (see for instance 
\cite{Matarrese:1984zw,Garriga:1999vw})
\be
P = X^{\frac{1+w}{2w}} \;, \quad w\neq 0 \;, \label{lw}
\ee
up to a proportionality constant which is irrelevant for the classical theory.
In this case, the speed of sound in eq.~(\ref{cs2}) reduces to $c_s^2 = w$.
As an example, one can consider a radiation fluid with equation of
state $w=1/3$. In this case the Lagrangian (\ref{lw}) reduces to
\be
{\cal L} = X^2 =  (-\partial_\mu \phi \partial^\mu \phi)^2 \;.
\ee

The inclusion of gravity is completely straightforward. The equation
of motion (\ref{eom2}) becomes
\be
\partial_\mu  [ \sqrt{-g} P'(X)\partial^\mu \phi ] =0\;. \label{eom3}
\ee
In an expanding FLRW background, the homogeneous solution satisfies now
\be
n = P' \dot \phi  \propto a^{-3} \;.
\ee
Indeed, for a constant $w$ this is the standard redshift of the energy
density,
\be
\rho \propto \dot \phi^{\frac{1+w}{w}} \propto a^{-3(1+w)} \;.
\ee

To summarize, we are able to describe the dynamics of one or more
fluids coupled with gravity, within a Lagrangian formalism. This can
be employed to study cosmological perturbations in the presence of
barotropic fluids. Note that the Lagrangian (\ref{lphi}) is a
particular case of the so called $k-$essence/$k-$inflation scenarios
\cite{ArmendarizPicon:1999rj,ArmendarizPicon:2000ah}. For this class
of Lagrangians,
cosmological perturbations have been
extensively studied starting from \cite{Garriga:1999vw}
and more recently extended to 
second \cite{Seery:2005wm,Chen:2006nt} and third order
\cite{Arroja:2008ga} to study the
non-Gaussianities produced during inflation. In this paper 
these calculations can be
reinterpreted as describing the nonlinearities of a fluid.

In the following we will be interested in the study of dark matter
perturbations, which would correspond to the dust case
$w=0$. Obviously the zero pressure limit must be taken with care, as
from eq.~(\ref{fq}) the Lagrangian strictly vanishes for a pressureless
fluid.

\section{Background and linear dynamics}
\label{bd}

Following the discussion of the last section, the dynamics of a
perfect fluid coupled with gravity can be described by the action
\begin{equation}
S = \frac{1}{2}\int \rd^4 x \sqrt{-g}[R + 2P(X)] \;,
\label{gravaction}
\end{equation}
where we chose units such that $M^{-2}_P \equiv 8 \pi G_N=1$. In particular
we are interested in the case $p = w \rho$ with constant $w$, when
$P$ is given by eq.~(\ref{lw}).

We assume a flat FLRW metric
$ d s^2 = - d t^2 + a^2(t) d \vec x^2$. Friedmann equations read
\bea
H^2&=&{1\over 3}\(2 X P' -P\) \;,\\
2\dot{H}&+& 3 H^2=-P \;,
\eea
where $H \equiv \dot{a}/a$ is the Hubble expansion parameter. It is
useful to define
\be
\epsilon\equiv -{\dot{H}\over H^2}={3\over2}(1+w) \;.
\label{epsilon}
\ee
Note that, although the
notation is inspired by the inflationary case,
we are not assuming that $\epsilon$ is small.\footnote{To simplify the
comparison with the literature we give here the parameters used in
refs.~\cite{Seery:2005wm,Chen:2006nt} for the $w = $ const case:
\be
\eta\equiv{\dot\epsilon\over \epsilon H}=0\;, \qquad
u\equiv 1-{1\over c^2_s}=1-{1\over w}\;, \qquad s\equiv {1\over
  H}{\dot{c_s}\over c_s}=0\;,
\ee
and
\be
\Sigma = X P'+ 2 X^2 P''= {H^2 \epsilon \over w}\;, \qquad
\lambda =  X^2 P'' + {2\over 3} X^3 P'''=-{H^2 \epsilon\over 6
  w}\(1-{1\over w}\) \;.
\ee}

Following Maldacena \cite{Maldacena:2002vr}, the study of
perturbations can be done using the ADM splitting of the metric \cite{ADM},
\begin{equation}
ds^2 = -N^2dt^2 + h_{ij}(dx^i + N^idt)(dx^j + N^jdt)\;.
\label{ADMmetric}
\end{equation}
The action (\ref{gravaction}) becomes
\begin{equation}
S = \frac{1}{2}\int \rd t\,\rd^3 x \sqrt{h}\left[N(R^{(3)} + 2P) + N^{-1}\left(E_{ij}E^{ij} - E^2\right)\right]\;,
\label{ADMaction}
\end{equation}
where $R^{(3)}$ is the curvature scalar computed from $h_{ij}$, $E \equiv E^i_{\phantom{i}i}$ and
\begin{equation}
E_{ij} \equiv \frac{1}{2}\left(\dot{h}_{ij} - \nabla_i N_j - \nabla_j N_i\right)\;.
\label{Edef}
\end{equation}
The covariant derivatives $\nabla_i$ are with respect to the 3-metric
$h_{ij}$ and all roman indeces $i,j, \ldots$ are raised and lowered
with this metric.

The action (\ref{gravaction}) describes 3 dynamical degrees of
freedom: one scalar mode for the fluid excitations and the 2 tensor
helicities of the gravity waves. In the ADM formalism, these degrees of freedom are contained
in the scalar field $\phi$ and in the 3-metric $h_{ij}$, while the
lapse $N$ and the shift $N_i$ are not dynamical. Following \cite{Maldacena:2002vr} we
choose the gauge
\begin{equation}
\begin{array}{c}
\delta\phi = 0, \qquad h_{ij} = a^2e^{2\zeta} \hat{h}_{ij} \;, \qquad \hat{h}_{ij} = \delta_{ij} + \gamma_{ij} + \frac{1}{2}\gamma_{il}\gamma_{lj}+\dots\;, \qquad  \\ \\
\mathrm{det}\;\hat{h} = 1 \;, \qquad \gamma_{ii}=0\;, \qquad
\partial_i\gamma_{ij}=0 \;,
\end{array}
\label{gaugeconditions}
\end{equation}
where $\zeta$ describes the scalar mode and $\gamma$ the tensor
ones. This gauge is comoving in the sense that the fluid 4-velocity is
everywhere orthogonal to the equal time surfaces\footnote{We follow
  the notation of \cite{Maldacena:2002vr} using $\zeta$ to denote the
  curvature perturbation in the comoving gauge. Often this quantity is
  instead called ${\cal R}$ reserving $\zeta$ to denote the curvature
perturbation in the uniform density gauge.}.

As the action does not contain time derivatives of $N$ and $N_i$,
these variables act as Lagrange multipliers, i.e.~their equations of
motion are algebraic constraints. These equations are the momentum and
Hamiltonian constraints of Einstein equations:
\begin{equation}
\nabla_i\left[N^{-1}\left(E^i_j - \delta^i_jE\right)\right] = 0 \;,
\label{momentumconstraint}
\end{equation}
\begin{equation}
R^{(3)} + 2P - 4X P' - \frac{1}{N^2}\left(E_{ij}E^{ij} -
  E^2\right) = 0 \;.
\label{energyconstraint}
\end{equation}

To warm up let us calculate the action for the scalar mode $\zeta$ at
second order, in order to find the linear equation of motion and the first
order metric. As there is no mixing between $\zeta$ and $\gamma$ at
second order in the action, in this section we can set $\gamma =0$.
To compute the second order action, we need to solve the constraints
(\ref{momentumconstraint},\ref{energyconstraint}) and plug their
solution for $N$ and $N_i$ back into the action
(\ref{ADMaction}). This needs to be done
at first order only, since the second-order solutions for $N$ and
$N^i$ will multiply $\delta {\cal L} /
\delta N$ or $\delta {\cal L} / \delta N^i$ at zeroth order, which vanish on
the background \cite{Maldacena:2002vr}.
To solve the momentum constraint (\ref{momentumconstraint}) we decompose $N_i$ as $N_i \equiv
\partial_i\psi + N_{Ti}$ where $\partial_i N_{Ti} = 0$. By defining
$N=1+ \delta N$, one finds, at first order,
\begin{equation}
\delta N = \frac{\dot{\zeta}}{H}\;, \qquad N_{T\,i}=0 \;.
\label{N1}
\end{equation}
Furthermore, one can find $\psi$ at first order by solving the energy
constraint\footnote{Here and in the following $\partial^2 \equiv \partial_i
  \partial_i$ and $\partial^{-2}$ is its inverse.} (\ref{energyconstraint}),
\begin{equation}
\psi = -\frac{\zeta}{H} +
\frac{a^2\epsilon}{w}\partial^{-2}\dot{\zeta} \;.
\label{psi1}
\end{equation}
Note that the last term of the previous equation contains $w$ at the
denominator so that one is not allowed to take the limit $w \to 0$ at
this stage.

Substituting the solutions (\ref{N1},\ref{psi1}) into the action
(\ref{ADMaction}) one obtains, after some integration by parts,
the second order action for $\zeta$,
\begin{equation}
S_2 = \int \rd t\,\rd^3x\, a^3\frac{\epsilon}{w} \left[\dot{\zeta}^2 -
  \frac{w}{a^2}(\partial \zeta)^2\right] \;.
\label{S2}
\end{equation}
The equation of motion derived from this action is thus
\begin{equation}
\ddot{\z} + 3H\dot{\z}-\frac{w}{a^2}\d^2\z = 0 \;.
\label{eom1}
\end{equation}

We are interested in the limit $w \to 0$. Setting $w=0$, the growing
solution of eq.~(\ref{eom1}) is simply a constant, $\zeta = \zeta_0$, where $\zeta_0$ is
the perturbation generated during inflation, which remains constant on
super-Hubble scales independently of the equation of
state\footnote{For a generic equation of state $p(\rho)$ the action for $\zeta$
  is
\be
S=\int \rd t \,\rd^3x \; a^3 \frac{\epsilon}{c_s^2} \left[ \dot \z^2 -
  \frac{c_s^2}{a^2} (\partial \zeta)^2 \right] \;,
\ee
where $c_s^2$ is given by eq.~(\ref{cs}). This gives 
the following equation of motion
\be
\ddot \z + \dot \z \frac{d}{dt} \ln \left(a^3 \frac{\epsilon}{c_s^2} \right)
  - \frac{c_s^2}{a^2} \partial^2 \zeta = 0\;.
\ee
From this equation we see that $\zeta$ is constant on large scales,
$k/aH \ll 1$, independently of the energy content.
This is not true, for instance, for the Newtonian potential.}. However, to
find the metric we need to plug the solution for $\zeta$ into
eqs.~(\ref{N1},\ref{psi1}) and for $\dot\zeta$ in eq.~(\ref{psi1}) we need to keep the leading order
correction in $w$ in eq.~(\ref{eom1}).

The general solution of this equation in Fourier space 
can be written as a linear combination of Bessel functions,
\be
\z_{\vec{k}}(\tau) = C_1\tau^{-\nu} J_\nu(k\tau\sqrt{w}) +
C_2\tau^{-\nu} Y_{\nu}(k\tau\sqrt{w}), \qquad \nu = \frac{3(1-w)}{2(1+3w)}\;,
\ee
where we have introduced the conformal time $\tau = \int dt/a$.
The growing mode of this solution goes as
\begin{equation}
\z = \z_0 +\frac{2 w}{5a^2H^2}\d^2\z_0 + \mathcal{O}(w^2)\;,
\label{z1}
\end{equation}
where $\zeta_0$ is the asymptotic value of $\zeta$ on super-Hubble
scales.
The time derivative of (\ref{z1}) is
\begin{equation}
\dot{\z} = \frac{2w}{5a^2 H}\d^2\z_0 + \mathcal{O}(w^2)\;.
\label{zdot1}
\end{equation}
This can be plugged into eqs.~(\ref{N1},\ref{psi1}) which yield, at
lowest order in $w\to 0$,
\be
\delta N=0\;, \qquad \psi = -\frac{2}{5}\frac{\z_0}{H}\;.
\ee
The metric at first order in perturbations is thus given by
\be
ds^2 = - dt^2 - \frac{4}{5H}\d_i\z_0 \; dt dx^i + a^2(1 + 2\z_0) d \vec x^2
\;.
\ee

Note that under a coordinate transformation
\be
t \to t  + \frac{2 \zeta_0}{5 H}\;, \qquad x^i \to x^i\;, \label{fogt}
\ee
this metric takes the Newtonian gauge form
\be
ds^2 = -(1+2 \Phi) dt^2 + a^2 (1-2 \Psi) d \vec x^2\;,
\ee
with
\be
\Phi=\Psi = -\frac{3}{5} \zeta_0 \;,
\ee
which is the well known linear relation between the Newtonian
potential and $\zeta_0$ in matter dominance.

\section{Second order perturbations}
\label{second}

\subsection{Scalar perturbations}
\label{sec:sm}

We are interested in deriving the third order action for $\zeta$. In
order to do that we follow the same procedure we did at linear order
and initially set $\gamma=0$ everywhere. Tensor modes generated at
second order by scalar perturbations will be considered in the next
section. To derive the third order action we
only need the solutions to the constraint equations at linear order,
i.e. eqs.~(\ref{N1},\ref{psi1}). This is because the second and third
order solutions would multiply the first and zeroth order constraint
equations, respectively \cite{Maldacena:2002vr}. We will be interested
in the second order solutions of the constraint equations only later,
when we will derive the explicit expression of the metric at second order.

It is straightforward to derive
\begin{eqnarray}
E_{ij}E^{ij} - E^2 & = & -6(H+\dot{\zeta})^2 + 4\frac{e^{-2\zeta}}{a^2}(H+\dot{\zeta})\left(\partial_iN_i + \partial_i\zeta N_i\right) \nonumber \\
& & + \frac{e^{-4 \zeta}}{a^4}\left[ \frac14 (\partial_i N_j
  +\partial_j N_i)^2 - (\partial_i N_j
  +\partial_j N_i)(N_i \partial_j \z + N_j \partial_i \z) -
  (\partial_i N_i)^2\right] 
\label{EE-E^2}
\end{eqnarray}
and
\begin{equation}
R^{(3)} = -2\frac{e^{-2\zeta}}{a^2}[2\partial^2\zeta
+\left(\partial\zeta\right)^2] \;.
\label{R}
\end{equation}

Expanding $P(X)$ with $X = \dot{\phi}^2 /N^2 $ and using the first
equality in (\ref{N1}) one obtains \cite{Seery:2005wm}
\begin{eqnarray}
P(X) & = & P + P'\left(-2\frac{\dot{\zeta}}{H} +
  3\frac{\dot{\zeta}^2}{H^2}
- 2 \,\delta N_2 - 4\frac{\dot{\zeta}^3}{H^3} + 6 \,\delta N_2\frac{\dot{\zeta}}{H}\right) X \nonumber \\
&&+\frac{1}{2}P''\left(4\frac{\dot{\zeta}^2}{H^2} + 8
  \,\delta N_2\frac{\dot{\zeta}}{H} - 12\frac{\dot{\zeta}^3}{H^3}\right)X^2 -
\frac{4}{3}P'''\frac{\dot{\zeta}^3}{H^3} X^3 \;,
\label{P}
\end{eqnarray}
where, on the right hand side, $X$, $P$ and its derivatives
are evaluated at zeroth order. The lapse has been split as $N=1+\delta
N_1+ \delta N_2$, where $\delta N_1 =
\dot\zeta/H$ is the first order contribution, eq.~(\ref{N1}), while
$\delta N_2$ is the second order one. 
As we discussed, terms
containing $\delta N_2$ will not appear in the action; however they will be
important to derive the metric.

Plugging the previous expressions into the action and using the
solutions to the constraint equation (\ref{N1})
one obtains, after several integrations by parts, the third
order action for $\z$,
\begin{eqnarray}
S_3 & = & \int \rd t\,\rd^3x\, a^3\left[\frac{3
    \epsilon}{w}\dot\z^2\zeta - \frac{\epsilon}{3 H w}(2+\frac1w)
  \dot{\zeta}^3 - \frac{\epsilon}{a^2}(\partial\zeta)^2\zeta  \right. \nonumber \\
& & \left.
- \frac{2}{a^4}\partial_i\zeta\partial_i\psi\partial^2\psi +
  \frac{1}{2 a^4}\left(\partial_i\partial_j\psi\partial_i\partial_j\psi
    - (\partial^2\psi)^2\right)(3\z -\frac{\dot{\zeta}}{H})\right]\;,
\label{S3}
\end{eqnarray}
where $\psi$ is given by eq.~(\ref{psi1}),
\be
\psi = -\frac{\zeta}{H} +
\frac{a^2\epsilon}{w}\partial^{-2}\dot{\zeta} \;.
\ee
This action describes the cosmological scalar perturbations of a
perfect fluid with constant equation of state $w$. Notice that it is
just a particular case of the general action derived in
refs.~\cite{Seery:2005wm,Chen:2006nt} in the context of generalized
inflationary models, although it describes here a quite different
physical system.

The action is simplified by a field redefinition, $\z_n = \z - f(\z)$, where
\cite{Chen:2006nt}
\begin{eqnarray}
f(\z) & = & \frac{1}{wH}\z\dot{\z} + \frac{1}{4a^2H^2}\left[-(\d_i\z)^2 + \d^{-2}\d_i\d_j(\d_i\z\d_j\z)\right] \nonumber \\
& & +\frac{\epsilon}{2 H^2 w}\left[\d_i\z\d_i\d^{-2}\dot{\z} -
  \d^{-2}\d_i\d_j(\d_i\z\d_j\d^{-2}\dot{\z})\right]\;.
\label{fieldredef}
\end{eqnarray}
As $f$ is quadratic in $\zeta$, the field redefinition does not change
the second order action. Note that the function $f(\zeta)$
 contains either spatial gradients or time
derivatives of $\zeta$. Thus, for any constant $w$, the field redefinition (\ref{fieldredef})
vanishes on super-Hubble scales, so that in this regime
$\z$ and $\z_n$ coincide.
In terms of $\z_n$ the action reads
\begin{eqnarray}
S_3 & = & \int \rd t\,\rd^3x\,a^3\frac{\epsilon}{w}\left[\frac{2}{3}
  \left( \frac1w -1 \right) \frac{\dot{\z}_n^3}{H} +
  \frac{3}{2} \left(3-\frac1w \right)\z_n\dot{\z}_n^2 + \frac{1}{2a^2} (5 +w) \z_n(\d_i\z_n)^2 \nonumber \right.\\
& &
\left.
  -\left(2-\frac{\epsilon}{2}\right)\frac{\epsilon}{w}\,\dot{\z}_n\d_i\z_n\d_i\d^{-2}\dot{\z}_n
+\frac{\epsilon^2}{4w}\d^2\z_n(\d_i\d^{-2}\dot{\z}_n)^2\right].
\label{S3w}
\end{eqnarray}

We are interested in a dust fluid $w=0$.
The equation of motion at second order derived from this action can be
solved perturbatively by plugging the first order solutions $\z = \z_n =
\z_0$ into the second order terms.
At lowest order in $w \to 0$ one finds
\begin{equation}
\label{3rdw0}
\ddot{\z}_n + 3H\dot{\zeta}_n =\frac{1}{a^2}F_0 \;,
\end{equation}
where
\be
F_0 = - \z_0 \d^2\z_0 - \frac{5}{16}(\d_i\z_0)^2 -
\frac{3}{8}\d_i\d^{-2}(\d^2\z_0 \d_i\z_0)\;. \label{F0}
\ee
The right hand side of eq.~(\ref{3rdw0}) is negligible at very early
times,  when all the modes
are well out of the Hubble radius, $k/aH \ll1$, so that $\zeta_n=$
const is a solution. As in this regime
$\z$ and $\z_n$ coincide as discussed, this corresponds to
the well-known fact that $\z$ is conserved on super-Hubble scales, even nonlinearly.
In this regime the Fourier modes of $\zeta$ are decoupled and equal to
the initial condition set by inflation, $\zeta = \zeta_0$ for  $k/aH
\ll1$.
Solving the equation above and going back to $\z$ using
(\ref{fieldredef}) yields
\be
\z = \z_0 - \frac{1}{5a^2H^2}\d^{-2}\d_i\d_j(\d_i\z_0\d_j\z_0) \;.
\label{zeta_sec_evol}
\ee
This expression gives $\z$ at second order during matter
dominance, as a function of its initial condition $\zeta_0$. We
postpone its discussion to the end of this section and we proceed
to conclude the calculation of the second order metric: we need
to solve for $N$ and $N_i$ at second order.

As at first order, to solve for the whole metric we
need to keep terms of order $w$ in the second order evolution
equation which now reads
\begin{equation}
\ddot{\zeta}_n + 3H\dot{\zeta}_n - \frac{w}{a^2}\d^2\z_n
=\frac{1}{a^2}(F_0 + wF_1) + \frac{w}{a^4 H^2} G \;,
\label{eom}
\end{equation}
where
\begin{eqnarray}
F_1 &=& -5\z_0 \d^2\z_0 + \frac{1}{8}(\d_i\z_0)^2 + \frac{3}{4}\d_i\d^{-2}(\d^2\z_0\d_i\z_0)\;, \label{F1}\\
G &=& - \frac{21}{25}(\d^2\z_0)^2 - \frac{2}{5}\z_0\d^4\z_0 + \frac{1}{20}\d_i\z_0\d_i\d^2\z_0 \nonumber \\
& & +\frac{9}{200}\d^2(\d_i\z_0)^2 - \frac{3}{50}\d_i\d^{-2}(\d_i\z_0\d^4\z_0)\;. \label{G}
\end{eqnarray}
To obtain the $w$ corrections to the right-hand side we used the
linear evolution of $\zeta$ at order $w^2$:
\begin{equation}
\dot{\z} = \frac{2w}{5a^2H}\d^2\z_0 - \frac{6w^2}{25a^2H}\d^2\z_0 + \frac{4w^2}{35a^4H^3}\d^4\z_0 + \mathcal{O}(w^3).
\label{zdot1(2)}
\end{equation}

Solving eq.~(\ref{eom}) and writing the solution in terms of $\z$
using the field redefinition (\ref{fieldredef}) one finds
\begin{eqnarray}
\z & = & \z_0 - \frac{1}{5a^2H^2}\d^{-2}\d_i\d_j(\d_i\z_0\d_j\z_0)  \nonumber \\
& & + \frac{w}{5a^2H^2}\[2 \d^2\z_0 -4\z_0\d^2\z_0
+\frac{14}{5}(\d_i\z_0)^2
+\frac{18}{5}\d_i\d^{-2}(\d^2\z_0\d_i\z_0) \right. \nonumber \\
& & \phantom{-\frac{w}{5a^4H^4}} -\frac{1}{7a^2H^2} \left( \frac{27}{20}(\d^2\z_0)^2
+\frac{51}{10}\d_i\z_0\d_i\d^2\z_0
+
\frac{4}{5}(\d_i\d_j\z_0)^2 \right.  \nonumber \\
& & \phantom{-\frac{w}{5a^4H^4}} \left.
  \left.+\frac{1}{2}\d_i\d^{-2}(\d_i\z_0\d^4\z_0)
    +\frac{1}{5}\d^{-2}\d_i(\d_i\d_j\z_0\d_j\d^2\z_0) \right)\] +
  \mathcal{O}(w^2) \;.
\label{z}
\end{eqnarray}

Now that we have a solution for $\z$ including ${\cal{O}}(w)$
corrections, we can proceed to solve the
constraints to get the final second order metric.  The momentum constraint is a vector equation; taking its transverse part with the projector $\delta_{ij}-\partial_i\partial_j/\partial^2$ we obtain an equation for the transverse part of the shift vector $N_{T i}$ ($\partial_i N_{T  i} =0$),
\be
\label{transshift} N_{T  i} = -\frac45 \frac{1}{H}
\partial^{-2}\left[\partial_i \zeta_0 \partial^2\zeta_0 -
\partial^{-2}\partial_i\partial_k (\partial_k\zeta_0
\partial^2\zeta_0)\right] \;.
\ee
Notice that the transverse part of the shift vanished at first order, see eq.~(\ref{N1}).
The longitudinal part of the momentum constraint gives an equation for the lapse function at second order,
\begin{eqnarray}
\delta N_2 & = & \frac{w}{5a^2H^2}\[(\d\z_0)^2 - 4 \z_0 \d^2 \z_0 \] \nonumber \\
& - & \frac{2w}{175a^4H^4}\[3(\d^2\z_0)^2 + 14 \d_i\z_0\d_i\d^2\z_0 +
4(\d_i\d_j\z_0)^2 \]  + \mathcal{O}(w^2)\;.
\label{N2}
\end{eqnarray}
Although the lapse perturbation vanishes in the limit $w \to 0$, its
expression at order $w$ is necessary to solve the energy
constraint, similarly to what happened at first order. This gives the
second order correction to the shift function
\begin{eqnarray}
\psi_2 & = & \frac{1}{5H}\d^{-2}\Big[(\d_i\z_0)^2 -
3\d^{-2}\d_i\d_j(\d_i\z_0\d_j\z_0)  \nonumber \\
& & \phantom{\frac{1}{5H}\d^{-2}} + \frac{4}{5a^2H^2} \Big( \frac{3}{7}(\d^2\z_0)^2 +
  \d_i\z_0\d_i\d^2\z_0+\frac{4}{7}(\d_i\d_j\z_0)^2\Big) \Big].
\label{psi2}
\end{eqnarray}

Plugging in the ADM metric (\ref{ADMmetric}) the first order results
(\ref{N1}), (\ref{psi1}) and the second order ones (\ref{z}),
(\ref{transshift}), (\ref{N2}) and (\ref{psi2}),
we finally obtain the second order metric for $w=0$,
\begin{eqnarray}
g_{00} & = & -1 + \frac{4}{25a^2H^2}(\d_i\z_0)^2\;, \label{g00}\\
g_{0i} & = & -\frac{1}{5H}  \d_i \Big[ 2\z_0  - \d^{-2} (\d_j\z_0)^2+3\d^{-4}\d_j\d_k(\d_j\z_0\d_k\z_0) \nonumber \\
& & \phantom{-\frac{1}{5H}  \d_i}- \frac{4}{5a^2H^2}\d^{-2} \Big( \frac{3}{7}(\d^2\z_0)^2 +
\d_i\z_0\d_i\d^2\z_0+\frac{4}{7}(\d_i\d_j\z_0)^2\Big) \Big] \nonumber \\
& & -\frac45 \frac{1}{H} \partial^{-2}\left[\partial_i \zeta_0
\partial^2\zeta_0 - \partial^{-2}\partial_i\partial_k
(\partial_k\zeta_0 \partial^2\zeta_0)\right] \;,
\label{g0i}\\
g_{ij} & = & a^2\[1+2\z_0+2\z_0^2-\frac{2}{5a^2H^2}\d^{-2}\d_k\d_l(\d_k\z_0\d_l\z_0)\]\delta_{ij} + a^2\gamma_{ij}\;. \label{gij}
\end{eqnarray}

Before moving to the calculation of the tensor contribution
$\gamma$, let us discuss eq.~(\ref{zeta_sec_evol}) which describes
the second order evolution of $\zeta$. In Fourier space this
equation reads
\begin{equation}
\z_{\vec k} = \z_{0 \vec k} + \frac{1}{5a^2H^2} \int \!
\frac{d^3q}{(2\pi)^3} \; \frac{(\vec k \cdot \vec q)(k^2 -\vec k
\cdot \vec q)}{k^2} \; \z_{0\vec q} \; \z_{0\vec k-\vec q} \;.
\label{zeta_sec_evol_four}
\end{equation}
It is important to stress that the kernel inside the integral does
not vanish in the limit $k \to 0$. Surprisingly, this implies that
a very long wavelength mode well outside the Hubble radius evolves
in the presence of short wavelength perturbations: $\zeta$ is {\em
not} conserved on super-Hubble scales as its second order
contribution in the matter dominated era grows like the scale
factor $a\propto 1/a^2H^2$. However, this is not in contrast with
the literature on the conservation of $\zeta$
\cite{Salopek:1990jq}, as $\zeta$ is conserved when all the modes
are out of the Hubble radius, so that the second term on the right
hand side of eqs.~(\ref{zeta_sec_evol}) and
(\ref{zeta_sec_evol_four}) can be neglected.

The fact that $\zeta$ is not conserved in our case is not so
surprising after all. Indeed, at nonlinear order one can define an
infinite set of variables which coincide with $\zeta$, and which
are thus conserved, when all the gradients can be neglected, i.e.~out
of the Hubble radius. 
However, when sub-Hubble modes are considered, if one of these
variables remains constant, the others will not in general. In
fact, one can trivially define a variable which is conserved on
all scales, i.e.,
\begin{equation}
\tilde \z = \z + \frac{1}{5a^2H^2}\d^{-2}\d_i\d_j(\d_i\z\d_j\z)
\;,
\end{equation}
even though this has not a particular geometrical meaning and we
do not expect it to remain conserved when a different equation of
state is considered. In order to discuss the physical significance
of the variation of $\zeta$ one needs to compute at second order
the relationship with a measurable quantity, like Cosmic
Microwave Background temperature fluctuations. This goes beyond
the scope of this article.

\subsection{Gravitational waves}
\label{sec:gw}

So far we did not discuss gravitational waves. At linear order
tensor and scalar modes are decoupled, so that gravitational waves can
be completely neglected in the limit where the primordial contribution
generated by inflation is very small. Beyond the linear
approximation scalar and tensor modes mix, so that it is not
consistent to set $\gamma = 0$. In the following we are going to
assume that the amplitude of primordial tensor modes is very
small: gravitational waves will only be generated by the scalar modes through couplings of the form $\gamma\zeta\zeta$.

To study the generation of gravitational waves we need the quadratic action for
$\gamma$ and the cubic terms of the form $\gamma\zeta\zeta$. The
constraint equations will not enter neither in the derivation of the
action nor in obtaining the expression of the second order metric. Indeed $\gamma$ does
not appear at first order in the constraint equations and quadratic
terms are negligible as $\gamma = {\cal{O}}(\zeta^2)$.

Let us start with the quadratic action for $\gamma$.
Making use of the expressions
\begin{equation}
E_{ij}E^{ij} - E^2 \supset \frac{1}{4}\dot{\gamma}_{ij}\dot{\gamma}_{ij}
\end{equation}
and
\begin{equation}
R^{(3)} \supset \frac{1}{a^2}\left(\gamma_{ij}\d^2\gamma_{ij} +\frac{3}{4}(\d_i\gamma_{jk})^2 - \frac{1}{2}\d_i\gamma_{kj}\d_k\gamma_{ij}\right)\;,
\end{equation}
after integration by parts this yields
\begin{equation}
S_{\gamma\gamma}=\frac{1}{8}\int\rd
t\,\rd^3x\,a^3\(\dot{\gamma}_{ij}\dot{\gamma}_{ij} -
\frac{1}{a^2}\d_k \gamma_{ij}\d_k\gamma_{ij} \)\;.
\label{gg}
\end{equation}
For the $\z\z\gamma$ terms we use the expressions
\begin{equation}
R^{(3)} \supset \frac{4}{a^2}\gamma_{ij}\d_i\d_j\z - \frac{10}{a^2}\gamma_{ij}\z\d_i\d_j\z
\end{equation}
and
\begin{equation}
e^{3\z}N^{-1}a^3\(E_{ij}E^{ij}-E^2\) \supset
-a\dot{\gamma}_{ij}\(3\z-\frac{\dot{\z}}{H}\)\d_i\d_j\psi +
\frac{1}{a}\d_k\gamma_{ij}\d_i\d_j\psi\d_k\psi \;,
\end{equation}
and we obtain
\begin{eqnarray}
S_{\gamma\z\z} & = & \int\rd t\,\rd^3x\,a^3\left[-\frac{2}{Ha^2}\gamma_{ij}\d_i\dot{\z}\d_j\z - \frac{1}{a^2}\gamma_{ij}\d_i\z\d_j\z  \right. \nonumber \\
& & \left. \phantom{\int\rd t\,\rd^3x\,a^3}
  -\frac{1}{2a^2}\left(3\z-\frac{\dot{\z}}{H}\right)\dot{\gamma}_{ij}\d_i\d_j\psi + \frac{1}{2a^4}\d_l\gamma_{ij}\d_i\d_j\psi\d_l\psi\right] \;.
\label{ssg}
\end{eqnarray}
This part of the action is the same as the one derived by Maldacena \cite{Maldacena:2002vr}
for a scalar field with standard kinetic term. This is not surprising
as these couplings do not depend on $P(X)$ in (\ref{ADMaction}).

The equation of motion for $\gamma$ can be obtained by varying the actions
(\ref{gg}) and (\ref{ssg}) with respect to $\gamma_{ij}$. Here there
are no subtleties in the limit $w \to 0$ so that one can drop terms
containing $\dot\zeta$. 
Notice that $\gamma_{ij}$ is constrained to satisfy the
transverse and traceless conditions $\partial_i \gamma_{ij}=0$ and
$\gamma_{ii}=0$. As such, the $\gamma$'s in the action above must be
understood as projected by the transverse and traceless projector, i.e.,
\begin{equation}
\label{projector}
P_{ij \, kl}^{\rm TT}={1\over 2} \(P_{ik} P_{jl} + P_{jk} P_{il}- P_{ij} P_{kl}\)\;,
\end{equation}
where $P_{ij}$ is a symmetric transverse projector given by
\begin{equation}
P_{ij}\equiv \delta_{ij} -{\partial_i\partial_j\over \partial^2} \;.
\end{equation}
Thus, the evolution equation reads
\begin{equation}
\ddot\gamma_{ij}+ 3 H \dot\gamma_{ij}-{\partial^2\over
  a^2}\gamma_{ij}=   P_{ij \, kl}^{\rm TT} \left[ {2\over a^2}  \d_k\z_0 \d_l\z_0 +
  {4\over 25 a^4 H^2}  \partial^2 \left(\d_k\z_0 \d_l \z_0 \right) \right] \;,
\end{equation}
where we have simplified the term in the square braket on the  right hand side, taking into account
that we are interested only in its transverse component.
The solution of this equation, with the appropriate initial condition
$\gamma_{ij}=0$, reads
\be
\gamma_{ij}=-\frac{4}{5}
\[ 9\( \frac{1}{3} - \frac{j_1(k\tau)}{k \tau} \right) \partial^{-2}
+\frac{1}{5a^2H^2} \]
P_{ij \, kl}^{\rm TT} \left( \d_k\z_0\d_l\z_0 \right)\; ,
\label{gammasol}
\ee
where the spherical Bessel function $j_1(x)$ can be written in terms
of trigonometric functions as $j_1(x)=\sin(x)/x^2-\cos(x)/x$.

\section{Transforming to Poisson gauge}\label{poisson}

In this section we want to write our second order metric in conformal Poisson
gauge, i.e. in the form
\begin{equation}
\rd s^2 = a^2(\tau)\left\{-(1+2\Phi_P)\rd\tau^2+ 2 \omega_{P \,i}^\perp d\tau dx^i +
  \[(1-2\Psi_P)\delta_{ij} + \gamma_{P\, ij}\]\rd x^i \rd x^j\right\},
\label{poissonmetric}
\end{equation}
where $\omega_{P \,i}^\perp$ is transverse, $\partial_i \omega^\perp_{P\, i} =0$, and $\gamma_P$ is transverse and traceless.
Going to this gauge will enable us to compare our results with those obtained in
\cite{Matarrese:1997ay,Bartolo:2005kv}.

For the second order gauge transformation we use the notation of
\cite{Matarrese:1997ay} (apart from exchanging the name of the scalar potentials $\Phi$ and $\Psi$) and we write our metric as
\begin{equation}
\rd s^2 =
a^2(\tau)\left\{-(1+2\Phi_{\zeta})\rd\tau^2
+2 (\omega^\perp_{\zeta \,i} + \partial_i \omega_\zeta)
  \rd\tau\rd x^i + \[(1-2\Psi_\zeta )\delta_{ij} +
  \gamma_{ ij}\]\rd x^i \rd x^j\right\} \;,
\label{pmetric}
\end{equation}
where
\begin{eqnarray}
\Phi_\zeta & = & -\frac{2}{25a^2H^2}(\d_i\z_0)^2\;, \label{Phi}\\
\omega_\zeta & = & -\frac{2}{5aH}\z_0 + 
\frac{1}{5aH}\d^{-2}\[(\d_i\z_0)^2-3\d^{-2}\d_i\d_j(\d_i\z_0\d_j\z_0)\] 
\nonumber \\
& & + \frac{4}{25a^3H^3}\d^{-2}\[\frac{3}{7}(\d^2\z_0)^2 
+ \d_i\z_0\d_i\d^2\z_0+\frac{4}{7}(\d_i\d_j\z_0)^2\]\;, \label{omega}\\
\omega^\perp_{\zeta\,i} & =& -\frac{4}{5 a H} \partial^{-2}
\left[\partial_i \zeta_0 \partial^2\zeta_0 -
\partial^{-2}\partial_i\partial_j (\partial_j\zeta_0 \partial^2\zeta_0)\right] \;,\label{trans}\\
\Psi_\zeta & = & -\z_0 -\z_0^2 +\frac{1}{5a^2H^2}\d^{-2}\d_i\d_j(\d_i\z_0\d_j\z_0) \label{Psi} \;,
\label{gammaij}
\end{eqnarray}
and $\gamma_{ij}$ is given by eq.~(\ref{gammasol}).
The gauge transformation up to second order is, in conformal
coordinates $x^\mu = (\tau, x^i)$,
\begin{equation}
x^\mu \to x^\mu - \xi_1^\mu - \xi^\mu_2 + \frac{1}{2} \xi_1^\nu
\xi_1^\mu{}_{,\nu}
\end{equation}
where at each order $\xi^0=\alpha$ and $\xi^i=\partial^i \beta +
d^i$ with the vector $d^i$ transverse, $\partial_i d_i=0$. The
first order piece is fixed by eq.~(\ref{fogt}), i.e. $\alpha_1= -2
\zeta_0/(5aH)$ and $\beta_1=0=d_{1 \; i}$. The parameters of the gauge
transformation at second order can be obtained  from the second order
components of the metric
\be
 \omega_{P\;i}^\perp + \partial_i \omega_P = \omega_{\zeta\;i}^\perp +
 \partial_i \omega_\zeta + \frac{\alpha_1}{2} \partial_i\left[ 2 (\alpha_1' + 2aH
  \alpha_1 )  - \alpha_1' -4aH \alpha_1\right] - \frac12 \alpha_1' \partial_i
\alpha_1 - \d_i \alpha_2 + \d_i \beta_2'  + d'_{2\;i} \;,
\label{omegaP}
\ee
and
\be
\gamma_{P\;ij} = \gamma_{ij}+ \partial_{i} \alpha_1 \partial_{j}
\alpha_1 - \frac13
  \delta_{ij} (\partial_k \alpha_1)^2 + 2 \left( \partial_i \d_j -
  \frac13 \delta_{ij} \d^2 \right) \beta_2 + \partial_i d_{2\;j} + \partial_j d_{2\;i}\;.
\ee
In the previous equations we used that $\omega_\zeta = -2
\zeta_0/(5aH) = \alpha_1$ at first order.
Imposing that $\gamma_P$ is transverse
($\chi_P^\parallel=\chi_P^\perp=0$ in the notation of
\cite{Matarrese:1997ay}) allows to solve for $\beta_2$ and $d_{2 \; i}$:
\begin{eqnarray}
\beta_2 &=& - \frac{3}{25a^2H^2} \d^{-2} \left[\d^{-2} \d_i\d_j(\d_i\z_0\d_j
  \z_0) - \frac13 (\d_i \z_0)^2 \right]\;, \label{beta2} \\
d_{2\;i} & = & -\frac{4}{25 a^2 H^2} \partial^{-2}
\left[\partial^2\zeta_0 \partial_i\zeta_0 -
\partial^{-2}\partial_i\partial_j (\partial^2\zeta_0
\partial_j\zeta_0)\right] \;. \label{d2}
\end{eqnarray}
Imposing $\omega_P=0$ in eq.~(\ref{omegaP}) gives
\be
\alpha_1+ \alpha_2 = \omega_\zeta + \beta_2' \;. \label{alpha2}
\ee

At second order, the other components of the metric transform 
as \cite{Matarrese:1997ay}
\begin{eqnarray}
\Phi_P &=& \Phi_\zeta + \frac{\alpha_1}{2} \left(\alpha_1'' + 5 aH
  \alpha_1'+\frac{3}{2} a^2 H^2 \alpha_1  \right) + \alpha_1'^2 + \alpha_2'
+ aH\alpha_2\;, \\
\Psi_P &=& \Psi_\zeta + \frac{\alpha_1}{2} \left[ - \frac32 a^2 H^2
  \alpha_1 - aH \alpha_1' +2 \left( \Psi_\zeta' + 2aH \Psi_\zeta \right)
    \right] -\frac16 (\partial_i \alpha_1)^2 - aH \alpha_2 - \frac13
    \partial^2 \beta_2 \;,
\end{eqnarray}
which, with the conditions (\ref{beta2}), (\ref{d2}) and (\ref{alpha2}) above,
give
\begin{eqnarray}
{\Phi}_P & = & - \frac35 \z_0+
\frac{9}{25} \left[ \z_0^2 + \d^{-2}(\d_j\z_0)^2
  -3\d^{-4}\d_i\d_j(\d_i\z_0\d_j\z_0) \right] \nonumber \\
& &
+\frac{6}{175a^2H^2}\d^{-2}\[2(\d_i\d_j\z_0)^2
+5(\d^2\z_0)^2+7\d_i\z_0\d_i\d^2\z_0\]
\;, \\
{\Psi}_P & = & - \frac35 \z_0 -\frac{9}{25} \left[\z_0^2
+\frac{2}{3}\d^{-2}(\d_i\z_0)^2- 2\d^{-4}\d_i\d_j(\d_i\z_0\d_j\z_0) \right]
\nonumber  \\
& &
+\frac{6}{175a^2H^2}\d^{-2}\[2(\d_i\d_j\z_0)^2+5(\d^2\z_0)^2+7\d_i\z_0\d_i\d^2\z_0\]
\;, \\
\omega^\perp_{P \; i} &=& -\frac{24}{25 a H} \partial^{-2} 
\left[\partial^2\zeta_0 \partial_i
\zeta_0 - \partial^{-2}\partial_i\partial_j (\partial^2\zeta_0
\partial_j\zeta_0)\right]   \;, \label{giusta} \\
{\gamma}_{P \; ij} & = & -\frac{36}{5}
\left( \frac{1}{3} - \frac{j_1(k\tau)}{k \tau} \right) \partial^{-2}
P_{ij \, kl}^{\rm TT} \left( \d_k\z_0\d_l\z_0 \right)\; .
\end{eqnarray}
The transverse traceless projector above can be expanded to give 
\begin{equation}
P_{ij \, kl}^{\rm TT} \left( \d_k\z_0\d_l\z_0 \right) = - \d^{-2}
\left[ \d^2 \Theta_0 \delta_{ij} + \d_i \d_j \Theta_0 +2 (\d^2 \zeta_0
  \d_i \d_j \zeta_0 - \d_i \d_k \z_0 \d_j \d_k \z_0) \right]\;,
\end{equation}
with
\begin{equation}
\Theta_0 = - \frac12 \d^{-2} \left[ (\d^2 \z_0)^2 -(\d_i\d_j\z_0)^2 
\right]\;.
\end{equation}

This result can be compared with ref.~\cite{Bartolo:2005kv} by taking
into account that $\zeta_0 = -(5/3) \Phi$ at linear order and on
super-Hubble scales. The resulting metrics coincide up to a typo 
in eq.~(A.29) of \cite{Bartolo:2005kv}.\footnote{To obtain the correct result the right hand side of 
eq.~(A.29) of \cite{Bartolo:2005kv} should be multiplied 
 by a factor $-2$. Note that in their
notation a factor of $1/2$ multiplies all the second order
contributions, 
i.e., $\omega_{i} \equiv \omega_{1i}+\frac{1}{2} \omega_{2i}$ while in
our notation there is no factor of $1/2$.}
Notice that in order to perform the calculation in Poisson gauge, one has
to supplement the second order equations obtained in
\cite{Matarrese:1997ay} by the proper second order matching with
the initial condition provided by $\zeta$ \cite{Bartolo:2005kv}.
In our calculation this matching is already taken into account by
the fact that we are always using the conserved quantity $\zeta$.

\section{Discussion}
\label{conclusions}

In this article we have developed a new approach to nonlinear
cosmological perturbations, based on the equivalence between
cosmological perfect fluids and derivatively coupled scalar
fields. In this approach, perturbations can be studied using an ADM
perturbative expansion of the scalar field action, similarly to
what is done for the study of non-Gaussianities during inflation.

Using this method, we have calculated the second order metric
during matter domination in the comoving gauge
(\ref{gaugeconditions}), where the scalar fluctuations are
described by the comoving curvature perturbation $\zeta$. The
metric, which is the main result of the paper, is given by
\begin{eqnarray}
g_{00} & = & -1 + \frac{4}{25a^2H^2}(\d_i\z_0)^2\;, \label{g002}\\
g_{0i} & = & -\frac{1}{5H}  \d_i \Big[ 2\z_0  - \d^{-2} (\d_j\z_0)^2+3\d^{-4}\d_j\d_k(\d_j\z_0\d_k\z_0) \nonumber \\
& & \phantom{-\frac{1}{5H}  \d_i}- \frac{4}{5a^2H^2}\d^{-2} \Big(
\frac{3}{7}(\d^2\z_0)^2 +
\d_i\z_0\d_i\d^2\z_0+\frac{4}{7}(\d_i\d_j\z_0)^2\Big) \Big] \nonumber \\
& & -\frac45 \frac{1}{H} \partial^{-2}\left[\partial_i \zeta_0
\partial^2\zeta_0 - \partial^{-2}\partial_i\partial_j
(\partial_j\zeta_0 \partial^2\zeta_0)\right] \;,
\label{g0i2}\\
g_{ij} & = &
a^2 \exp[2 \zeta(t)]
\delta_{ij}
+ a^2\gamma_{ij}\;, \label{gij2}
\end{eqnarray}
where 
\be
\zeta(t) = \z_0-\frac{1}{5a^2H^2}\d^{-2}\d_k\d_l(\d_k\z_0\d_l\z_0) \;,
\label{zeta_finn}
\ee
$\zeta_0$ is the perturbation generated during inflation,
conserved on super-Hubble scales, and where $\gamma_{ij}$ is a
traceless transverse tensor, given by
\be
\gamma_{ij}=-\frac{4}{5}
\[ 9\( \frac{1}{3} - \frac{j_1(k\tau)}{k \tau} \right) \partial^{-2}
+\frac{1}{5a^2H^2} \] P_{ij \, kl}^{\rm TT} \left(
\d_k\z_0\d_l\z_0 \right)\; . \label{gammasol2}
\ee

Note that at early times, when all the modes are out of the Hubble
radius so that we can neglect gradients, the above metric is just
\be
ds^2 =-dt^2 + a^2(t) e^{2 \zeta_0(\vec x)} d \vec x^2 \;.
\ee
This is the form of the metric in comoving gauge after all the modes
have left the Hubble radius during inflation. This metric is known
to remain the same during the subsequent evolution of the Universe
until the modes re-enter the Hubble radius. This shows that our
metric nicely matches the initial conditions set by inflation.

As shown by eq.~(\ref{zeta_finn})
above, we have found that $\zeta$ is not conserved at second order during the
matter dominated era. Indeed, super-Hubble modes can evolve under
the presence of short sub-Hubble perturbations. However, we have
concluded that a discussion on the physical significance of this
variation cannot be done without the computation of the relation
between $\zeta$ and observable quantities, which is beyond the
scope of this article.

Our method can be generalized to compute the metric during a
cosmological era dominated by a fluid with constant equation of
state $w\neq 0$, like during radiation dominance, by choosing the
appropriate $w$ in the linear evolution equation for $\zeta$,
eq.~(\ref{eom1}), and in the third order action (\ref{S3w}).
The only difference is that in this case $\zeta$ is not constant
inside the Hubble (see eq.~(\ref{eom1})), so that the second order 
solution will involve an integral over conformal time.
Furthermore, one can generalize our method to study the evolution
of perturbations in a Universe filled with several coupled fluids,
by introducing a correspondent number of derivatively coupled scalar
fields in the action, with possible couplings between them. As a
future study, this method could be used to study the generation of
gravitational waves from second order perturbations during
radiation and matter dominated era \cite{Ananda:2006af,Baumann:2007zm}.

\section*{Acknowledgments}
We thank Francis Bernardeau, Chris Clarkson, Guido D'Amico, Ruth Durrer, David
Langlois, Sabino Matarrese, Alberto Nicolis, Cyril Pitrou, 
Riccardo Rattazzi and Jean-Philippe Uzan for useful discussions.

\end{document}